\documentclass[12pt,preprint]{aastex}

\slugcomment{ApJ Letters; Accepted for publication July 21, 2009}

\def\eq#1{\begin{equation} #1 \end{equation}}
\def\about  {\hbox{$\sim$}}

\def\E#1{\hbox{$10^{#1}$}}

\def\LEdd   {\hbox{$L_{\rm Edd}$}}
\def\Mo     {\hbox{$M_{\odot}$}}
\def\erg   {\hbox{erg\,s$^{-1}$}}
\def\kms   {\hbox{km\,s$^{-1}$}}
\def\Rd     {\hbox{$R_{\rm d}$}}
\def\Mout   {\hbox{$\dot M_{\rm out}$}}
\def\Macc   {\hbox{$\dot M_{\rm acc}$}}
\def\IBLR   {\hbox{$I_{\rm BLR}$}}
\def\ITOR   {\hbox{$I_{\rm TOR}$}}

 \shorttitle{BLR Disappearance in LLAGNs}
 \shortauthors{Elitzur \& Ho}

\begin{document}

\title{ON THE DISAPPEARANCE OF THE BROAD-LINE REGION IN LOW-LUMINOSITY AGNs}

\author{Moshe Elitzur\altaffilmark{1}
        and Luis C. Ho\altaffilmark{2}}

\altaffiltext{1}{Department of Physics and Astronomy, University of Kentucky,
                Lexington, KY 40506-0055; moshe@pa.uky.edu}
\altaffiltext{2}{The Observatories of the Carnegie Institution of Washington,
                813 Santa Barbara Street, Pasadena, CA 91101; lho@ociw.edu}

\begin{abstract}

The disk-wind scenario for the broad-line region (BLR) and toroidal obscuration
in active galactic nuclei predicts the disappearance of the BLR at low
luminosities. In accordance with the model predictions, data from a nearly
complete sample of nearby AGNs show that the BLR disappears at luminosities
lower than $5\times\E{39}\,(M/10^7\Mo)^{2/3}$\,\erg, where $M$ is the black
hole mass. The radiative efficiency of accretion onto the black hole is $\la
\E{-3}$ for these sources, indicating that their accretion is
advection-dominated.

\end{abstract}


\section{INTRODUCTION}

According to the unification model for active galactic nuclei (AGNs), the only
difference between type~1 (broad-line) and type~2 (narrow-line) sources is the
observer's orientation with respect to the toroidal dusty obscuration.
Directions with clear sight of the central engine and the broad-line region
(BLR) yield type~1 sources, while those blocked by the torus from direct view
of the BLR  result in type~2 objects, where the existence of the hidden BLR is
revealed only in polarized light \citep[e.g.,][]{Ski93}. However, in spite of
the considerable success of the unification scheme there is now clear
evidence, recently summarized by \cite{Ho08}, that the BLR is actually
missing, and not just hidden, in many low-luminosity AGNs (LLAGNs). These
sources have been named ``pure" \citep{Tran01,Tran03} or ``true"
\citep{Laor03} type~2 AGNs. Why does the BLR disappear in LLAGNs?

In another important recent development, the obscuring torus is found to be a
smooth continuation of the BLR, not a separate entity. Primary evidence comes
from the \cite{Suganuma06} discovery that the BLR extends outward until the
inner boundary of the dusty torus, in agreement with the \cite{Netzer_Laor}
proposal that the BLR outer boundary is set by dust sublimation. Additionally,
\cite{Risaliti02} find that X-ray variability caused by passage of absorbing
clouds across the line-of-sight has a continuous behavior across the time
scales generated by the motions of dust-free BLR clouds close to the AGN and
the more distant, dusty torus clouds. Together, these observations show that
the BLR and the torus are simply the inner and outer regions, respectively, of
a single, continuous cloud distribution. Their different radiative signatures
merely reflect the change in cloud composition across the dust sublimation
radius \Rd. The inner clouds are dust-free, and their gas is directly exposed
to the AGN ionizing continuum; therefore, the gas is atomic and ionized,
producing the broad emission lines. The outer clouds are dusty, and so their
gas is shielded from the ionizing radiation, and the atomic line emission is
quenched. Instead, these clouds are molecular and dusty, obscuring the
ultraviolet/optical emission from the inner regions and emitting infrared
radiation.  Within this framework, the BLR occupies $r < \Rd$ while the torus
is simply the region at $r > \Rd$, and a more appropriate designation for it is
the ``toroidal obscuration region'' \citep[TOR;][]{Elitzur07, AGN2}.

The BLR/TOR structure arises naturally in the disk-wind scenario, first
proposed by \cite{Emmering92}. In this model, the two classes of clouds simply
correspond to different regions of a clumpy wind coming off the accretion disk
rotating around the black hole \citep[see][and references
therein]{Elitzur_Shlosman}. As the clouds rise away from the disk they expand
and lose their column density, limiting the vertical scope of broad-line
emission and dust obscuration and emission, resulting in a toroidal geometry
for both the BLR and the TOR. Although a theory of clumpy disk winds in AGNs is
far from full development, an immediate consequence of this scenario is the
prediction that the TOR and BLR disappear at low bolometric luminosities
\citep[i.e., low accretion rates;][]{Elitzur_Shlosman, Elitzur08}. The reason
is that, as the mass accretion rate decreases, the mass outflow rate of a disk
wind with fixed radial column decreases more slowly and thus cannot be
sustained below a certain accretion limit. This unavoidable conclusion follows
from simple considerations of mass conservation. However, in the absence of a
complete theory for the outflow dynamics, the actual limit and its detailed
dependence on the AGN parameters remain undetermined. Presently, the only
practical approach is to attempt to extract clues on these unknowns from
analysis of LLAGN data. This is the approach taken here.

\section{The BLR Disappearance}

Denote by $L_{45} = L/\E{45}\ \erg$ the AGN bolometric luminosity and by $M_7 =
M/\E7 \Mo$ the black hole mass. From mass conservation, disk outflow can
support the column densities required by the BLR and TOR only when the
luminosity obeys
\eq{\label{eq:bound1}
    L_{45} > C\, M_7^{2/3}
}
\citep{Elitzur_Shlosman, Elitzur08}. The derivation of this result is repeated
below in \S \ref{sec:Mdot}, where we analyze in detail the assumptions and
provide the expression for the numerical coefficient $C$ in terms of system
properties other than $L$ and $M$. Replacing the mass with the Eddington
luminosity, $\LEdd = 1.3\times \E{45}\,M_7\,\erg$, produces the luminosity
lower bound in terms of Eddington ratio
\eq{\label{eq:bound2}
    L_{45} > C_2 \left({\LEdd\over L}\right)^2,
}
where $C_2 = 0.59\,C^3$. In the absence of a detailed theoretical model for the
disk wind, our ignorance about the dynamics is contained in the numerical
coefficient $C$. The only way to determine whether $C$ may contain additional,
indirect dependence on $L$ or $M$ is to compare the bounds in eqs.\
\ref{eq:bound1} and \ref{eq:bound2} with observations.

We use data selected from the Palomar spectroscopic survey of 486 nearby
galaxies \citep{Ho97a,Ho97b}, the most sensitive and complete sample of LLAGNs
available.  A critical new development is that the majority of the Palomar
galaxies now have reliable estimates of their nuclear luminosities and black
hole masses \citep{Ho09a,Ho09b,Ho+09}. We consider all galaxies classified as
Seyferts, low-ionization nuclear emission-line regions (LINERs), and transition
objects, which, as argued in \cite{Ho08,Ho09a}, are variants of LLAGNs that
define a sequence of varying accretion rates.  We also include absorption-line
nuclei, objects with undetected optical line emission that are good candidates
of hosting inactive black holes. High-resolution, hard-X-ray measurements
provide the most robust measure of the accretion luminosity in AGNs, especially
for weak sources.  Of the 277 Palomar galaxies meeting our classification
criteria, 166 (60\%) have suitable nuclear X-ray (2--10 keV) fluxes or upper
limits thereof \citep{Ho09a}. As discussed in \cite{Ho09a}, these X-ray
measurements can be converted to bolometric luminosities with an accuracy of
$\sim 0.3$ dex. Importantly, all but one of the galaxies has reliable
measurements of the central stellar velocity dispersion \citep{Ho+09}, from
which the black hole mass can be estimated to within $\sim 0.3$ dex using the
well-established relation between black hole mass and bulge stellar velocity
dispersion \citep{Tremaine02}.

Figure~1 shows the distributions of data points in the $L - M$ and $L -
L/\LEdd$ planes. The Palomar LLAGNs are coded according to spectral class and
presence or absence of broad H$\alpha$ emission \citep{Ho97b}.  For comparison,
we also include the sample of low-redshift luminous Seyfert 1 nuclei and
quasars studied by \cite{Greene_Ho}.  It is apparent that, depending on either
the black hole mass or Eddington ratio, type~1 sources cease to exist below a
certain luminosity. (The only exception is the LINER 1 galaxy NGC 2787, whose
luminosity of \E{39.98}\,\erg\ is 0.45 dex below the drawn boundary line.)
We interpret this critical luminosity to be the threshold below which the BLR
disappears in the disk-wind model.  The solid line in each panel represents the
bound with the slope taken from either eq.\ \ref{eq:bound1} or \ref{eq:bound2},
as appropriate. The line intercepts were adjusted to coincide with the observed
type~1$-$type~2 boundary, yielding $C = 4.7\times\E{-6}$, i.e., the BLR
existence requires $L \ga 5\times\E{39}\,\erg$ .

\section{The Disk-Wind Constraint}
\label{sec:Mdot}

Consider an outflow coming off an annular segment of a disk. Denote by $n(R)$
the density and $v_z(R)$ the vertical component of outflow velocity at axial
radius $R$ on the disk surface. Then the outflow mass loss rate is $\Mout =
2\pi m \int n v_z RdR$, where $m$ is the mean atomic mass. Since the dust
sublimation radius \Rd\ sets a characteristic scale for both the BLR and TOR,
it is convenient to introduce the scaled radius $y = R/\Rd$ and the
dimensionless velocity profile $u = v_z(R)/v_z(\Rd)$. Introduce also $N_R =
\int n(R)dR$, the overall column density along a radius vector on the disk
surface. We can describe the radial density variation by the dimensionless
profile $\eta = n(R)\Rd/N_R$, normalized from $\int\eta(y)dy = 1$. Then
\eq{\label{eq:basic}
    \Mout = 2\pi m N_R \Rd v_z(\Rd) I\,,
}
where
\[I = \int \eta(y)u(y)\,ydy\]
is a numerical factor of order unity. Written in terms of the radial column
density $N_R$, this basic expression provides the mass outflow rate of any disk
wind, whether smooth or clumpy; in the latter case $v_z$ is the cloud liftoff
velocity and the profile $\eta$ describes the variation of number of clouds per
unit radial length.

The dust sublimation radius in AGNs is $\Rd \simeq 0.4 L_{45}^{1/2}$\,pc
\citep{AGN2}. The initial outflow velocity, $v_z$, is roughly the dispersion
velocity of material in the disk --- an outflow is established when the ordered
motion velocity becomes comparable to that of the local random motions. We now
make the assumption that this velocity is some fraction $f_1$ of the local
Keplerian velocity, $v_z(\Rd) = f_1 (GM/\Rd)^{1/2}$. Maser observations of the
nuclear disk in NGC~3079 show that the velocity dispersion is \about\ 14 \kms\
in a small region of strong emission where the Keplerian velocity is 110 \kms\
\citep{Kondratko05}.  This observation suggests that $f_1$ is probably of order
\about\ 0.1. With this assumption, the mass outflow rate becomes
\eq{\label{eq:M_Out}
    \Mout = 0.07f_1 I N_{22}\, M_7^{1/2}L_{45}^{1/4}
                                                 \quad \Mo\,{\rm yr}^{-1},
}
where $N_{22} = N_R/\E{22}\,\rm cm^{-2}$. This expression applies both for the
BLR and TOR portions of the disk wind.

In steady state, \Mout\ cannot exceed \Macc, the mass accretion rate into the
BLR/TOR region. A fraction $f_2$ of this accreted mass finds its way to the
black hole and is converted to the AGN bolometric luminosity with radiative
efficiency $\eta$ so that $L = f_2\eta\Macc c^2$. Therefore, $\Macc = 0.02
L_{45}/(\eta f_2)\, \Mo\,{\rm yr}^{-1}$ and
\eq{
    {\Mout\over\Macc} = \epsilon
                {M_7^{1/2}\over L_{45}^{3/4}},
}
where $\epsilon = 3.5\eta f_1 f_2 I N_{22}$. Steady-state mass conservation
requires $\Mout/\Macc < 1$, yielding eq.\ \ref{eq:bound1} with $C =
\epsilon^{4/3}$. The observational result $C = 4.7\times\E{-6}$ implies $\eta
f_1 f_2 I N_{22} = 3\times\E{-5}$. From photoionization modeling of broad-line
emission, the BLR requires a minimal column density of \about\ \E{22} cm$^{-2}$
\citep{Netzer90}, and so $N_{22} \ga 1$. We can reasonably take $I \simeq 1$
and $f_1 \simeq 0.1$, yielding $f_2\eta \simeq 3\times\E{-4}$; since the
factors $f_2$ and $\eta$ do not enter separately, only their product is
constrained. A rough estimate of $f_2$ can be obtained from self-similar
hydromagnetic disk-wind models. The wind applies a back torque on the
underlying disk because each magnetic field line can be considered rigid from
its footpoint $R$ up to the Alfv\'en radius $R_{\rm A}$. Angular momentum
conservation implies that $\gamma \equiv (R/R_{\rm A})^2$ is roughly the ratio
of the mass fluxes for outflow and accretion, or $\gamma \sim
\Mout(R)/\Macc(R)$.  This ratio is found to be either independent of $R$
\citep{Emmering92} or weakly dependent on it \citep{Pelletier92}, with a value
of $\gamma \simeq 0.1-0.3$.  The fraction of \Macc\ that reaches the black hole
is $1 - \gamma = f_2$, therefore $f_2 \ga 0.7$ and $\eta \simeq 4\times\E{-4}$.
Such a low radiative efficiency places the BLR disappearance in the domain of
radiatively inefficient or advection-dominated accretion flow solutions
\citetext{for recent reviews, see \citealp{Narayan02}; \citealp{Yuan07}}. A
``standard" accretion efficiency $\eta \simeq 0.1$ would require virtually the
entire mass accreted into the BLR/TOR to be carried away by the wind, with only
a fraction $f_2 \simeq 3\times\E{-3}$ reaching the black hole. This seems
unlikely.

\section{Discussion}

Quenching of the BLR/TOR at low accretion rates is an unavoidable consequence
of the disk-wind scenario. The reason is that the mass outflow rate is set by
the radial column density $N_R$ and by the scales of the local radius and
Keplerian velocity (eq.\ \ref{eq:basic}). Since the BLR/TOR boundary is set by
a fixed value of the radiative flux, the radial scale increases with luminosity
as $L^{1/2}$ and the velocity scale decreases as $L^{-1/4}$. Keeping $N_R$
constant, as appropriate for both the BLR and TOR, the mass outflow rate varies
with $L$ as $L^{1/4}$, more slowly than the linear variation of the accretion
rate. Disk-wind mass conservation implies that the minimal column required by
an observable BLR cannot be sustained below a certain luminosity. The data
verify that this is indeed the case and indicate that the transition to
``true'' type 2 AGNs occurs in the regime of radiatively inefficient accretion.
This adds support to independent indicators that LLAGNs accrete in the
advection-dominated mode \citep[][and references therein]{Ho09a}.

Bounds similar to eq.\ \ref{eq:bound1} were derived by \cite{Nicastro00} and
\cite{Laor03}. Both concluded independently that the BLR should disappear when
the luminosity drops below some $L_{\rm min} = CM^\alpha$, where $C$ and
$\alpha$ are constants. Replacing $M$ with the Eddington ratio, this condition
produces significantly different results depending on the value of $\alpha$:
when $\alpha < 1$ it implies a lower limit $L > C_2\,(\LEdd/L)^{\alpha\over 1 -
\alpha}$, where $C_2$ is derived from $C$, but when $\alpha > 1$ the result is
an {\em upper} bound $L < C_2\,(L/\LEdd)^{\alpha\over\alpha - 1}$; in between,
$\alpha = 1$ implies that the BLR exists only above some fixed Eddington
ratio.\footnote{As $\alpha$ increases, the bound is rotating clockwise in the
$\log L - \log (L/\LEdd)$ plane, passing through vertical at $\alpha = 1$, with
BLR sources located ahead of the rotating boundary.} Nicastro suggests a disk
outflow origin for the BLR, as here, and uses the disk/corona model of
\cite{Witt97} to derive $\alpha = 7/8$, practically indistinguishable from the
$\alpha = 2/3$ found here. However, Nicastro's bound translates into $L >
C_2(\LEdd/L)^7$, a much steeper boundary than indicated by the data in the $L -
L/\LEdd$ plane (Figure 1). Laor suggests the existence of an upper limit to BLR
velocities, which leads to $\alpha = 2$. Because $\alpha > 1$ in this case, it
yields $L < C_2(L/\LEdd)^2$. This bound is orthogonal to the one plotted in
panel (b) of Figure 1 and all AGN with BLR should lie {\em below} it, a
prediction that seems in clear conflict with the data. As these examples show,
examining the data in the $L - L/\LEdd$ plane amplifies the differences between
models; while eq.\ \ref{eq:bound2} describes adequately the data in Figure
1(b), the other models do not.

The derivation of the BLR/TOR disappearance presented here is more general than
the original one in \cite{Elitzur_Shlosman}. Equation \ref{eq:M_Out} holds
irrespective of the wind clumpiness and applies equally to the BLR and TOR
portions of the disk outflow, with $N_{22}$ the radial column density through
each pertinent segment. In both regions, $N_{22} \sim 1$ can be considered a
reasonable lower limit for the BLR to generate line emission and the TOR to
produce substantial obscuration. The only significant difference between the
two cases enters from the factor $I$ (eq.\ \ref{eq:basic}): denoting the inner
and outer radii of the disk wind by $R_{\rm in}$ and $R_{\rm out} $,
respectively, the integration in \IBLR\ extends from $R_{\rm in}/\Rd$ to 1,
while in \ITOR\ it is from 1 to $R_{\rm out}/\Rd$. We can reasonably assume
Keplerian behavior for the velocity profile, i.e., $u = y^{-1/2}$, and
parametrize the density profile with a power law, $\eta \propto y^{-q}$, so
that the integrand is proportional to $y^{1/2 - q}$. Depending on $q$,
different relations exist between \IBLR\ and \ITOR. The range $1 \le q \le 3/2$
yields $\ITOR/\IBLR \approx 1$, $q < 1$ produces $\ITOR/\IBLR \approx (R_{\rm
out}/\Rd)^{1/2} > 1$, while $q > 3/2$ results in $\ITOR/\IBLR \approx (R_{\rm
in}/\Rd)^{1/2} < 1$. The larger is $I$, the earlier does the quenching of the
outflow occur as the luminosity is decreasing. Therefore, depending on the
steepness of the radial density profile, the BLR and TOR may disappear either
together, when $\IBLR \sim \ITOR$, or in sequence: in AGNs with steep profiles
($q > 3/2$; $\IBLR > \ITOR$) the BLR will disappear first; in shallow ones ($q
< 1$; $\ITOR > \IBLR$) the TOR. Recent clumpy torus modeling of extended IR
data sets \citep{Mor09, Almeida09} suggests that $q$ can range anywhere from 0
to 3 in different sources, indicating that no single rule applies to all AGNs
--- sources with and without TOR can exist on either side of the BLR boundary.
In any given AGN, the TOR disappearance can be established from the lack of IR
emission at a level commensurate with the bolometric luminosity. The absence of
a TOR in M87 was demonstrated conclusively by \cite{Whys04}, who placed
stringent limits on the thermal IR emission, and further solidified by
\cite{Perlman07}, who found only a trace of thermal emission that is much
weaker than expected from an AGN torus and that can be attributed to
neighboring dust. Detailed studies of the IR emission from sources close to the
BLR boundary established in Figure 1 can be a powerful tool for probing the
density distribution around the black hole in LLAGNs and clarifying the nature
of the BLR/TOR disk outflow.

\acknowledgments

Insightful comments by Roberto Maiolino provided the impetus for this work. ME
acknowledges partial support by NSF (AST-0807417) and NASA (SSC-40095).

\newpage


\begin{thebibliography}{31}
\expandafter\ifx\csname natexlab\endcsname\relax\def\natexlab#1{#1}\fi

\bibitem[{{Antonucci}(1993)}]{Ski93}
{Antonucci}, R. 1993, \araa, 31, 473

\bibitem[{{Elitzur}(2007)}]{Elitzur07}
{Elitzur}, M. 2007,  in The Central Engine of Active Galactic
  Nuclei, ed. L.~C. {Ho} \& J.-W. {Wang} (San Francisco, CA: ASP), 415

\bibitem[{{Elitzur}(2008)}]{Elitzur08}
{Elitzur}, M. 2008, New Astron. Rev., 52, 274

\bibitem[{{Elitzur} \& {Shlosman}(2006)}]{Elitzur_Shlosman}
{Elitzur}, M., \& {Shlosman}, I. 2006, \apjl, 648, L101

\bibitem[{{Emmering} {et~al.}(1992){Emmering}, {Blandford}, \&
  {Shlosman}}]{Emmering92}
{Emmering}, R.~T., {Blandford}, R.~D., \& {Shlosman}, I. 1992, \apj, 385, 460

\bibitem[{{Greene} \& {Ho}(2007)}]{Greene_Ho}
{Greene}, J.~E., \& {Ho}, L.~C. 2007, \apj, 667, 131

\bibitem[{{Ho}(2008)}]{Ho08}
{Ho}, L.~C. 2008, \araa, 46, 475

\bibitem[{{Ho}(2009{\natexlab{a}})}]{Ho09a}
{Ho}, L.~C. 2009{\natexlab{a}}, \apj, in press

\bibitem[{{Ho}(2009{\natexlab{b}})}]{Ho09b}
{Ho}, L.~C. 2009{\natexlab{b}}, \apj, in press

\bibitem[{{Ho} {et~al.}(1997{\natexlab{a}}){Ho}, {Filippenko}, \&
  {Sargent}}]{Ho97a}
{Ho}, L.~C., {Filippenko}, A.~V., \& {Sargent}, W.~L.~W. 1997{\natexlab{a}},
  \apjs, 112, 315

\bibitem[{{Ho} {et~al.}(1997{\natexlab{b}}){Ho}, {Filippenko}, {Sargent}, \&
  {Peng}}]{Ho97b}
{Ho}, L.~C., {Filippenko}, A.~V., {Sargent}, W.~L.~W., \& {Peng}, C.~Y.
  1997{\natexlab{b}}, \apjs, 112, 391

\bibitem[{{Ho} {et~al.}(2009){Ho}, {Greene}, {Filippenko}, \&
  {Sargent}}]{Ho+09}
{Ho}, L.~C., {Greene}, J.~E., {Filippenko}, A.~V., \& {Sargent}, W.~L.~W. 2009,
  \apj, in press

\bibitem[{{Kondratko} {et~al.}(2005){Kondratko}, {Greenhill}, \&
  {Moran}}]{Kondratko05}
{Kondratko}, P.~T., {Greenhill}, L.~J., \& {Moran}, J.~M. 2005, \apj, 618, 618

\bibitem[{{Laor}(2003)}]{Laor03}
{Laor}, A. 2003, \apj, 590, 86

\bibitem[{{Mor} {et~al.}(2009){Mor}, {Netzer}, \& {Elitzur}}]{Mor09}
{Mor}, R., {Netzer}, H., \& {Elitzur}, M. 2009, \apj, submitted

\bibitem[{{Narayan}(2002)}]{Narayan02}
{Narayan}, R. 2002, in Lighthouses of the Universe: The Most Luminous Celestial
  Objects and Their Use for Cosmology, ed. M.~{Gilfanov}, R.~{Sunyaev}, \&
  E.~{Churazov} (Berlin: Springer), 405

\bibitem[{{Nenkova} {et~al.}(2008){Nenkova}, {Sirocky}, {Nikutta},
  {Ivezi{\'c}}, \& {Elitzur}}]{AGN2}
{Nenkova}, M., {Sirocky}, M.~M., {Nikutta}, R., {Ivezi{\'c}}, {\v Z}., \&
  {Elitzur}, M. 2008, \apj, 685, 160

\bibitem[{{Netzer}(1990)}]{Netzer90}
{Netzer}, H. 1990, in Active Galactic Nuclei, ed. T.~J.~L. {Courvoisier} \&
  M.~{Mayor} (Berlin: Springer), 57

\bibitem[{{Netzer} \& {Laor}(1993)}]{Netzer_Laor}
{Netzer}, H., \& {Laor}, A. 1993, \apjl, 404, L51

\bibitem[{{Nicastro}(2000)}]{Nicastro00}
{Nicastro}, F. 2000, \apjl, 530, L65

\bibitem[{{Pelletier} \& {Pudritz}(1992)}]{Pelletier92}
{Pelletier}, G., \& {Pudritz}, R.~E. 1992, \apj, 394, 117

\bibitem[{{Perlman} {et~al.}(2007){Perlman}, {Mason}, {Packham}, {Levenson},
  {Elitzur}, {Schaefer}, {Imanishi}, {Sparks}, \& {Radomski}}]{Perlman07}
{Perlman}, E.~S., {et~al.} 2007, \apj, 663, 808

\bibitem[{{Ramos Almeida} {et~al.}(2009){Ramos Almeida}, {Levenson}, {Rodriguez
  Espinosa}, {Alonso-Herrero}, {Asensio Ramos}, {Radomski}, {Packham},
  {Fisher}, \& {Telesco}}]{Almeida09}
{Ramos Almeida}, C., {et~al.} 2009, \apj, submitted

\bibitem[{{Risaliti} {et~al.}(2002){Risaliti}, {Elvis}, \&
  {Nicastro}}]{Risaliti02}
{Risaliti}, G., {Elvis}, M., \& {Nicastro}, F. 2002, \apj, 571, 234

\bibitem[{{Suganuma} {et~al.}(2006){Suganuma}, {Yoshii}, {Kobayashi},
  {Minezaki}, {Enya}, {Tomita}, {Aoki}, {Koshida}, \& {Peterson}}]{Suganuma06}
{Suganuma}, M., {et~al.} 2006, \apj, 639, 46

\bibitem[{{Tran}(2001)}]{Tran01}
{Tran}, H.~D. 2001, \apjl, 554, L19

\bibitem[{{Tran}(2003)}]{Tran03}
{Tran}, H.~D. 2003, \apj, 583, 632

\bibitem[{{Tremaine} {et~al.}(2002){Tremaine}, {Gebhardt}, {Bender}, {Bower},
  {Dressler}, {Faber}, {Filippenko}, {Green}, {Grillmair}, {Ho}, {Kormendy},
  {Lauer}, {Magorrian}, {Pinkney}, \& {Richstone}}]{Tremaine02}
{Tremaine}, S., {et~al.} 2002, \apj, 574, 740

\bibitem[{{Whysong} \& {Antonucci}(2004)}]{Whys04}
{Whysong}, D., \& {Antonucci}, R. 2004, \apj, 602, 116

\bibitem[{{Witt} {et~al.}(1997){Witt}, {Czerny}, \& {Zycki}}]{Witt97}
{Witt}, H.~J., {Czerny}, B., \& {\.{Z}ycki}, P.~T. 1997, \mnras, 286, 848

\bibitem[{{Yuan}(2007)}]{Yuan07}
{Yuan}, F. 2007, in The Central Engine of Active Galactic
  Nuclei, ed. L.~C. {Ho} \& J.-W. {Wang} (San Francisco, CA: ASP), 95

\end{thebibliography}



\begin{figure}

\includegraphics[width=\hsize,clip=true]{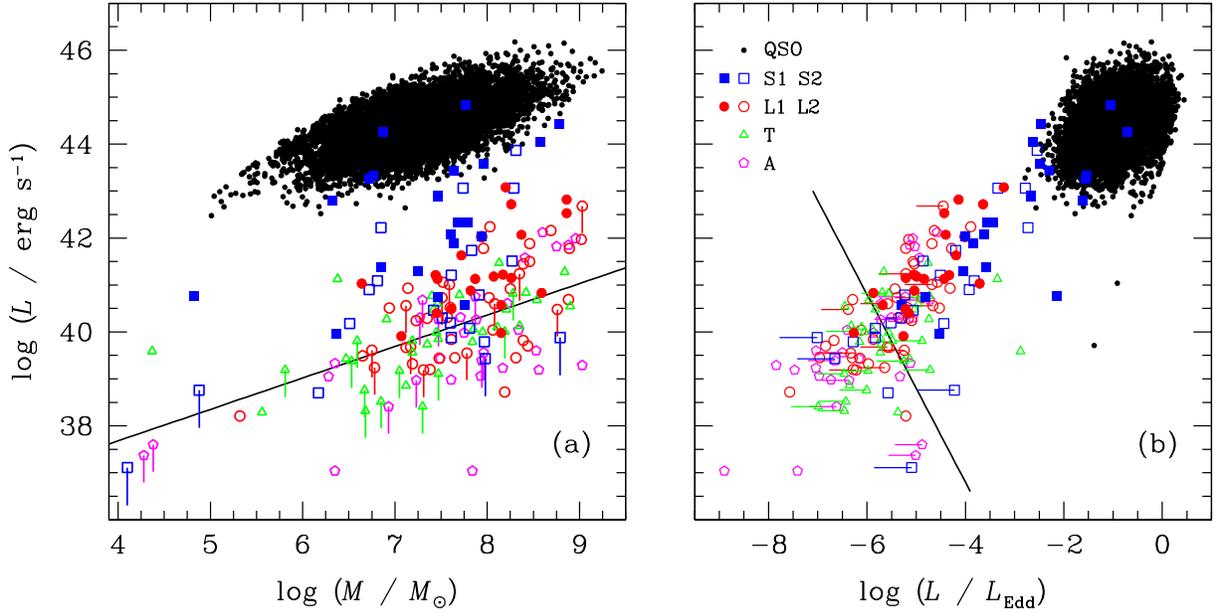}

\caption{Distribution of ({\it a}) black masses, $M$, and ({\it b}) Eddington
ratios, $L/L_{\rm Edd}$, vs.  bolometric luminosity, $L$, for objects separated
by spectral classification.  The bolometric luminosity is based on the 2--10
keV X-ray luminosity, as described in Ho (2009a).  The symbols are identified
in the legend.  Solid and open symbols indicate type~1 and type~2 sources,
respectively.  ``S'' = Seyferts, ``L'' = LINERs, ``T'' = transition objects,
and ``A'' = absorption-line nuclei.  Short line segments denote upper limits.
The objects marked as ``QSO'' refer to the sample of high-luminosity Seyfert 1
nuclei and quasars studied by Greene \& Ho (2007). The solid lines are
$\log L = 35 + \frac23\log M$ in panel ({\it a})  and $\log L = 28.8 -
2\log (L/\LEdd)$ in ({\it b}), corresponding to the theoretical predictions
given in eq.\ (1) and eq.\ (2), respectively.
}
\end{figure}

\end{document}